\documentclass[%
 aip,
 amsmath,amssymb,
preprint,%
]{revtex4-1}

\usepackage{graphicx}
\usepackage{dcolumn}
\usepackage{bm}

\usepackage[utf8]{inputenc}
\usepackage[T1]{fontenc}
\usepackage{mathptmx}
\usepackage{etoolbox}
 \usepackage{booktabs} 

\makeatletter
\def\@email#1#2{%
 \endgroup
 \patchcmd{\titleblock@produce}
  {\frontmatter@RRAPformat}
  {\frontmatter@RRAPformat{\produce@RRAP{*#1\href{mailto:#2}{#2}}}\frontmatter@RRAPformat}
  {}{}
}%
\makeatother
\begin{document}

\preprint{AIP/123-QED}

\title[Rod-climbing effect revisited]{The role of the second normal stress difference in rod-climbing effect}
\author{Rishabh V. More}
 \email{rishabh.more@monash.edu.}
\affiliation{ Department of Chemical and Biological Engineering, Monash University, Clayton, 3800, VIC, Australia}

\date{\today}

\begin{abstract}
The Weissenberg (rod-climbing) effect, i.e., the rise of a viscoelastic fluid along a thin rotating rod, has long served as a canonical demonstration of elasticity and normal-stress differences in complex fluids. The effect is most commonly attributed to the first normal stress difference $N_{1}$, which induces tensile hoop stresses that draw fluid upward along the rod. The second normal stress difference $N_{2}$, in contrast, is often presumed negligible or dynamically unimportant. However, many polymer solutions and industrial fluids, such as suspensions, exhibit $N_{2}$ of appreciable magnitude, and modern constitutive models predict that it can significantly modify free-surface stresses and thereby the climbing behaviour. In this work, we perform high-resolution axisymmetric simulations of the Linear Phan--Thien--Tanner (LPTT) model to systematically isolate the influence of $N_{2}$ on rod climbing. We show that increasing the magnitude of $N_{2}$ progressively weakens the climbing response and ultimately reverses it, producing rod-descending once the normal-stress ratio exceeds a critical value $\psi_{0}\approx 0.25$. Larger $N_{2}$ also destabilises the flow, promoting early onset (in terms of the rotation speed) of bubble formation, subcritical Hopf oscillations, and fully asymmetric three-dimensional motion that culminates in rupture. By mapping these regimes in the $(Wi,\psi_{0})$ parameter space, where $Wi$ is the Weissenberg number, we reconcile discrepancies among perturbation theory, experiments, and numerical simulations. These results establish $N_{2}$ as a crucial control parameter governing free-surface stability in viscoelastic liquids.

\end{abstract}


\maketitle

\section{Introduction}

The rod-climbing or Weissenberg effect was first described in detail by Weissenberg, who observed that a rotating rod draws a viscoelastic liquid upward along its surface \cite{weissenberg1947continuum, bird1987dynamics}. This striking phenomenon played a central role in establishing the existence of normal stresses in polymeric liquids, one of the four key rheological phenomena \cite{ewoldt2022designing} and subsequently motivated a sequence of theoretical developments in the 1970s. Joseph and coworkers developed a domain-perturbation framework for predicting free-surface deformation of a second-order fluid in rotating-rod geometries, culminating in a series of foundational analyses that derived the scaling of the climbing height at small rotation rates \cite{joseph1973free1,joseph1973free2,beavers1980free3}. Their work showed that the leading-order interface deformation varies quadratically with the rotation speed $\Omega$ and depends on a climbing constant 
\[
\hat{\beta}=0.5\,\Psi_{1,0}+2\,\Psi_{2,0},
\]
a combination of the zero–shear-rate normal-stress coefficients
\begin{equation}\label{eq:eq1}
 \lim_{\dot{\gamma}\to 0}\frac{N_1}{\dot{\gamma}^2}=\Psi_{1,0},
 \qquad 
 \lim_{\dot{\gamma}\to 0}\frac{N_2}{\dot{\gamma}^2}=\Psi_{2,0},
\end{equation}
where $N_1$ and $N_2$ denote the first and second normal stress differences, respectively and $\dot{\gamma}$ is the shear rate. At sufficiently small rotation rates, the sign and magnitude of $\hat{\beta}$ determine whether the free surface rises or descends. More precisely, the perturbation to the static meniscus height $h_{s}$ for a rod of radius $a$ rotating at angular speed $\Omega$ is given by \cite{joseph1966rodclimbing,more2023rod}
\begin{equation}\label{eq:eq2}
\begin{aligned}
    \Delta h(\Omega,a)
    &= h(\Omega,a,\alpha)-h_{s}(a,\alpha) \\
    &\approx \frac{a}{2(\Gamma \rho g)^{1/2}}
    \left[
        \frac{4\hat{\beta}}{4+\sqrt{Bo}}
        - \frac{\rho a^{2}}{2+\sqrt{Bo}}
    \right]\Omega^{2}
    + O(\Omega^{2}\alpha+\Omega^{4}),
\end{aligned}
\end{equation}
where $\Gamma$ is the surface tension, $\alpha$ is the contact angle, $\rho$ is the fluid density, and $Bo=\rho g a^{2}/\Gamma$ is the Bond number with $g$ being the acceleration due to gravity.

Although classical perturbation theory correctly identified that both $N_{1}$ and $N_{2}$ contribute to $\hat{\beta}$, it did not distinguish their separate dynamical roles. As a result, the mechanistic influence of $N_{2}$ remains poorly understood. Experiments by Beavers and Joseph \cite{beavers1975rotating} confirmed the quadratic scaling predicted by Eq.~\ref{eq:eq2}, yet they also revealed systematic deviations at moderate rotation rates where higher-order normal-stress interactions become significant. These discrepancies raised the possibility that $N_{2}$, long assumed negligible, might influence rod climbing more strongly than previously believed. Indeed, early rheological measurements demonstrated that $N_{2}$ may reach $10$–$30\%$ of $N_{1}$ in many polymer solutions \cite{magda1991second,magda1993rheology,magda1994concentrated}, while reptation theory predicts that the normal-stress ratio $\psi_0=-\Psi_{2,0}/\Psi_{1,0}$ lies between $1/7$ and $2/7$ depending on molecular alignment assumptions \cite{kimura1981polym}. On the other hand, $|N_2|$ can be $\gg |N_1|$ in dense \cite{boyer2011dense, more2020roughness} and fiber suspensions \cite{khan2023rheology}. 

The precise value of $\psi_0$ critically affects the magnitude and even the sign of the climbing response. If $\psi_{0}=1/4$, then the climbing constant $\hat{\beta}$ vanishes and Eq.~\ref{eq:eq2} predicts rod descending instead of climbing. Neglecting inertia yields the classical climbing condition $\hat{\beta}>0$, i.e.,
\[
\psi_{0}<0.25,
\]
for climbing to be observed \cite{beavers1975rotating, beavers1980free3}. However, this asymptotic inequality is incomplete because inertial effects can compete directly with elastic stresses at experimentally relevant rotation rates. Retaining inertia in Eq.~\ref{eq:eq2} leads to a more general small-$\Omega$ climbing condition \cite{more2023rod}:
\begin{equation}
\psi_{0} 
< \frac{1}{4}\left[
1 - 
\frac{4+\sqrt{Bo}}{2(2+\sqrt{Bo})}
\frac{\rho a^{2}}{\Psi_{1,0}}
\right],
\label{eq:climb_condition}
\end{equation}
which reveals that the transition between climbing and descending results from a competition between the elasticity encoded in $\Psi_{1,0}$, the normal-stress ratio $\psi_0$, and inertia quantified by the dimensionless parameter $\rho a^{2}/\Psi_{1,0}$ \cite{more2023rod}. This generalised criterion explains, for instance, why a weakly viscoelastic 0.3 wt.\,\% polyisobutylene (PIB) solution, which satisfies $\psi_{0}<0.25$, nevertheless undergoes rod descending due to dominant inertial effects \cite{more2023rod}. Conversely, highly elastic dilute Boger fluids may climb strongly despite $\psi_{0}$ values close to theoretical predictions $\psi_0 = 2/7$ from reptation theory \cite{magda1991second, magda1994concentrated}.

Rod climbing has also long been used to measure normal stresses. In the low-$\Omega$ limit, the climbing constant $\hat{\beta}$ can be used to infer combinations of $\Psi_{1,0}$ and $\Psi_{2,0}$ \cite{magda1991second, beavers1975rotating, hossain2024protorheology, maklad2021review}. More recent work revisited this idea in the context of modern torsional rheometry and showed that $\hat{\beta}$, combined with independent small-amplitude oscillatory shear (SAOS) measurements of $\Psi_{1,0}$, can be used to extract $\Psi_{2,0}$ \cite{more2023rod}. However, experimental challenges \cite{maklad2021review} in isolating $N_{2}$ have hindered systematic studies, leaving open fundamental questions regarding its role in rod climbing.

\begin{figure} \centering \includegraphics[width=0.5\linewidth]{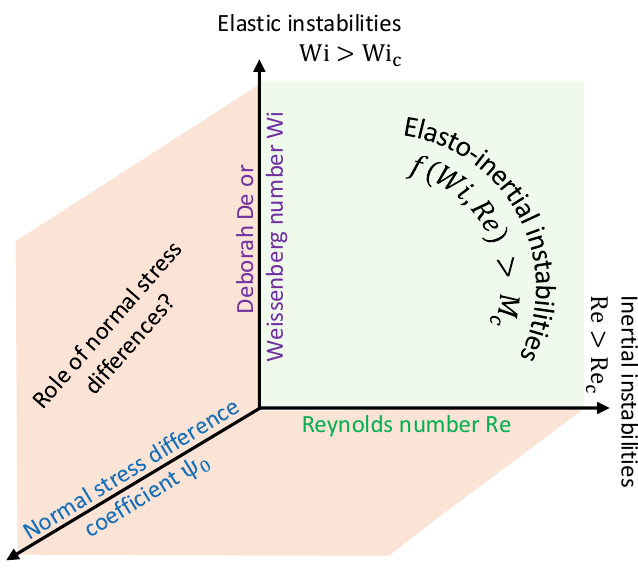} \caption{A three-dimensional representation of the parameter space accessed in the instabilities in complex flows of complex fluids. The Weissenberg-Reynolds number space has been extensively studied and it is well known that complex flows become unstable due to 1) Elastic instabilities if $Wi > Wi_c$ when $Re \ll 1$, or 2) Inertial instabilities if $Re > Re_c$ when $Wi \ll 1$, or 3) a combination $f(Wi, Re)>M_c$ in the intermediate limit \cite{datta2022perspectives, more2024elasto}. However, the role of the second normal stress difference remains incompletely understood and is represented as the third dimension of this parameter space. This third dimension is explored using the popular rod-climbing problem in this study.} \label{fig:fig1} \end{figure}

A further motivation for the present study stems from a modern reinterpretation of viscoelastic flow instabilities. A recent perspective article by Datta et al. \cite{datta2022perspectives} emphasises that complex-fluid instabilities span a broad three-dimensional parameter space defined by inertia, elasticity, and surface tension. Their analysis shows that examining instabilities solely in the Weissenberg-Reynolds number ($ Wi, Re$) plane obscures important effects arising from additional material functions, such as $\Psi_2$. Motivated by this viewpoint, we extend the classical stability landscape into a third axis corresponding to the normal-stress ratio $\psi_{0}=-N_{2}/N_{1}$, as illustrated in Fig.~\ref{fig:fig1}. This extension is especially relevant because normal-stress differences play a central role in the dynamics and instability of many free-surface and confined complex fluid flows \cite{maklad2021review, more2024elasto}.

Such instabilities have been documented extensively in rotating-rod experiments, which reveal nonlinear transitions including steady bubbles, periodic oscillations, and even chaotic free-surface motion \cite{beavers1975rotating,degen1998time,more2023rod}, none of which can be explained within the confines of second-order perturbation theory. Axisymmetric finite-element simulations by Luo (1999)\cite{luo1999numerical} further demonstrated that strongly nonlinear stress fields develop near the rod even at modest $Wi$, underscoring the need for fully coupled viscoelastic modelling. More recently, two-phase algorithms developed by Figueiredo and Oishi (2016)\cite{figueiredo2016two} have captured high-Weissenberg-number free-surface instabilities, including interface oscillations, for Oldroyd--B fluids ($\psi_{0}=0$). A recent transient analysis based on the Giesekus model showed that the second normal stress difference reduces the climbing height \cite{ruangkriengsin2025transient}; however, this model does not allow independent variation of $\psi_{0}$, and therefore the quantitative influence of the normal-stress ratio on the rod-climbing height remains unresolved. Despite these advances, the mechanisms by which $N_{2}$ modifies both the onset and character of rod-climbing instabilities continue to pose open questions.

In this context, the present work systematically investigates the effect of the second normal-stress difference on rod climbing within a constitutive model in which the normal-stress ratio $\psi_{0}$ can be varied independently. The Linear Phan–Thien–Tanner (LPTT) model provides such a framework and enables controlled exploration of the mechanistic influence of $N_{2}$ on steady and unsteady free-surface responses. By mapping the dependence of rod climbing on both $Wi$ and $\psi_{0}$, we identify the conditions under which \emph{viscoelastic} rod descending occurs, reconcile discrepancies between classical theory and experiments, and expand the landscape of viscoelastic-flow instabilities envisioned in Fig.~\ref{fig:fig1}.

\section{Governing Equations and Theoretical Framework}

\begin{figure}
    \centering
    \includegraphics[width=\linewidth]{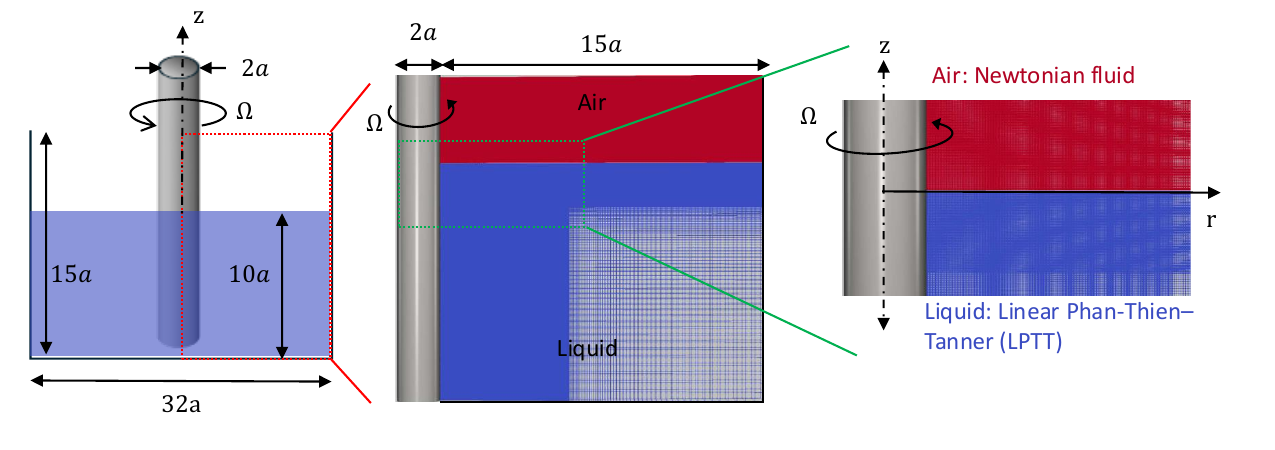}
    \caption{Problem set up. We compute the flow of a Linear Phan-Thien-Tanner viscoelastic liquid \cite{phan1978tanner} with varying second normal stress difference around a thin rod of radius $a=0.00635$ m rotating at a rotational speed of $\Omega \in [0 - 100]$ rad/s. The flow is assumed to be 2-D axisymmetric, and a non-uniform grid is used, finer near the rod and progressively coarser in the r-direction, shown at progressively higher zoom levels. No-slip boundary conditions are applied to all solid surfaces. Acceleration due to gravity $g=9.81$ m/s$^2$ acts in the negative $z$ direction.}
    \label{fig:fig2}
\end{figure}

The rod-climbing problem set up is illustrated in Fig.~\ref{fig:fig2}a. The liquid is assumed to be incompressible, with density $\rho$ and velocity field $\mathbf{u}=(u_r,u_\theta,u_z)$ in cylindrical coordinates $(r,\theta,z)$. The governing equations are the axisymmetric incompressible Navier–Stokes equations,
\begin{equation}
\nabla\cdot \mathbf{u} = 0,
\end{equation}
\begin{equation}
\rho\left(\frac{\partial \mathbf{u}}{\partial t}+\mathbf{u}\cdot\nabla\mathbf{u}\right)
= -\nabla p + \nabla\cdot\boldsymbol{\tau},
\end{equation}
where $p$ is the pressure and $\boldsymbol{\tau}$ is the stress tensor. Surface tension enters the free-surface boundary condition through the normal-stress jump,
\begin{equation}
\mathbf{n}\cdot \boldsymbol{\tau}\cdot \mathbf{n}  = \Gamma \kappa,
\end{equation}
where $\Gamma$ is surface tension and $\kappa$ the curvature of the interface. Tangential stress continuity is enforced as
\begin{equation}
\mathbf{t}\cdot \boldsymbol{\tau}\cdot \mathbf{n} = 0.
\end{equation}
Here $\mathbf{n}$ and $\mathbf{t}$ are the normal and tangential unit vectors at the interface, respectively. The stress is written as the sum of a Newtonian solvent contribution with a viscosity $\eta_s$ and a viscoelastic polymer contribution $\boldsymbol{\tau}_p$,
\begin{equation}
\boldsymbol{\tau} = 2\eta_s\mathbf{D} + \boldsymbol{\tau}_p,
\end{equation}
where $\mathbf{D}=\frac{1}{2}(\nabla\mathbf{u} + \nabla\mathbf{u}^T)$ is the rate-of-deformation tensor. We use the Linear Phan-Thien Tanner \cite{phan1978tanner} constitutive equation to model the polymeric contribution, which satisfies the following evolution equation
\begin{equation}
\lambda\overset{\triangledown}{\boldsymbol{\tau}_p} = 2\eta_p \mathbf{D} - f(\mathrm{tr}\boldsymbol{\tau}_p)\,\boldsymbol{\tau}_p - 2\psi_0 (\boldsymbol{\tau}_p \cdot \mathbf{D} + \mathbf{D} \cdot \boldsymbol{\tau}_p)
\end{equation}
with relaxation time $\lambda$, polymeric viscosity $\eta_p$, and the damping function with a constant $\epsilon$
\begin{equation}
f = 1 + \epsilon \frac{\lambda}{\eta_p}\mathrm{tr}(\boldsymbol{\tau}_p).
\end{equation}
The upper-convected derivative is
\[
\overset{\triangledown}{\boldsymbol{\tau}_p}
= \frac{\partial\boldsymbol{\tau}_p}{\partial t}
+ \mathbf{u}\cdot\nabla\boldsymbol{\tau}_p
- (\nabla\mathbf{u})^T\cdot\boldsymbol{\tau}_p
- \boldsymbol{\tau}_p\cdot(\nabla\mathbf{u}).
\]

For the LPTT model, the first and second normal stress differences in steady shear are given by
\[
N_1(\dot{\gamma}) = \tau_{\theta \theta} - \tau_{rr} = 2\lambda\eta_p \dot{\gamma}^2 \, f(\epsilon,\dot{\gamma}), 
\qquad
N_2(\dot{\gamma}) = \tau_{rr} - \tau_{zz} = -\psi_0 N_1(\dot{\gamma}),
\]
where $\psi_0 = -N_2/N_1$ is the normal-stress ratio. In this study, $\psi_0$ is varied directly, allowing independent adjustment of $N_2$ while keeping all other liquid properties constant for isolating the effect of $N_2$.

\section{Methodology}

The computational problem setup shown in Fig.~\ref{fig:fig2}b-c corresponds to a cylindrical reservoir containing a rotating rod of radius $a$. The rod rotates with angular velocity $\Omega$; gravity acts in the negative $z$ direction. Axisymmetry is assumed. The free surface is resolved using a geometric volume-of-fluid method, and the stress and velocity fields are solved on a nonuniform axisymmetric mesh refined near the rod where shear is the largest. The Reynolds number $Re = \rho \Omega a^2 / (\eta_s + \eta_P)$ is kept sufficiently small to isolate elastic effects. The Weissenberg number is defined as $ Wi = \lambda \Omega$. The normal-stress ratio $\psi_0$ is varied parametrically from $0$ to $0.25$. Time integration is performed until either a steady solution or a long-time periodic state emerges.

The numerical simulations are performed using the open-source 
finite-volume framework \textsc{OpenFOAM} in conjunction with the viscoelastic flow solver \textsc{RheoTool} \cite{pimenta2017stabilization}, which provides robust algorithms for differential constitutive models. The governing equations are discretised using second-order finite-volume schemes on a nonuniform axisymmetric mesh refined near the rotating rod. To ensure numerical stability at moderate and high Weissenberg numbers, the polymeric stresses are computed using the log-conformation representation $\boldsymbol{\Theta}$ \cite{pimenta2017stabilization, habla2014numerical}, in which the evolution equation is solved for the matrix logarithm of the conformation tensor rather than for the stress tensor $\boldsymbol{\tau}_P$ itself. This formulation maintains the positive definiteness of the conformation tensor and greatly enhances robustness in strongly elastic flows, consistent with earlier studies that have employed the method for viscoelastic instabilities \cite{fattal2004constitutive}. The polymeric stress tensor is reconstructed from the log-conformation tensor $\boldsymbol{\Theta}$ through a transformation \cite{habla2014numerical}. 
Extensive validations \cite{habla2014numerical} of \textsc{RheoTool} demonstrate its reliability for predicting viscoelastic flows with free surfaces and instability-driven dynamics, and therefore, it provides a suitable numerical platform for the present 
rod-climbing study.

The viscoelastic liquid considered in this study is modelled using the Linear Phan--Thien--Tanner (LPTT) constitutive equation with material parameters selected to match the rheology of a moderately elastic polymer solution. The density of the fluid is $\rho_{f}=890~\text{kg\,m}^{-3}$ and the surface tension at the free surface is $\Gamma=0.0309~\text{N\,m}^{-1}$. The total zero-shear viscosity is decomposed into a Newtonian solvent contribution $\eta_{s}=1.46~\text{Pa\,s}$ and a polymeric contribution $\eta_{p}=13.14~\text{Pa\,s}$, giving a viscosity ratio $\beta=\eta_{s}/(\eta_{s}+\eta_{p})$. The elastic relaxation time is fixed at $\lambda=0.018225~\text{s}$, and the LPTT parameter $\epsilon$ is set to $\epsilon=0$, corresponding to the linear version of the model. With this choice, the damping function simplifies to $f=1$, ensuring that nonlinearity in the polymer stress arises solely through the upper-convected derivative. 

These LPTT fluid parameters produce a viscoelastic fluid with a well-defined first normal stress difference and a tunable second normal stress difference through the normal-stress ratio $\psi_{0}$, allowing a direct investigation of how $N_{2}$ influences free-surface deformation and rod-climbing behaviour. In addition, this facilitates validation of the methodology (see Appendix A), with previous experimental and numerical results at these parameter values in the literature \cite{figueiredo2016two, luo1999numerical, habla2011numerical}. We vary $\psi_0$ from 0 to 0.25 to systematically explore the effect of the second normal stress difference on rod-climbing by keeping the first normal stress difference and other material properties constant as tabulated in Table~\ref{tab:LPTTparams}. Furthermore, the dimensionless ratio $\rho a^{2}/\Psi_{1,0} \approx 0.1$ indicates that elastic stresses dominate over inertial contributions in this regime.

\begin{table}[h!]
\centering
\caption{Material parameters used for the Linear Phan--Thien--Tanner (LPTT) liquid.}
\begin{tabular}{l c c}
\toprule
\textbf{Parameter} & \textbf{Symbol} & \textbf{Value} \\
\midrule
Density & $\rho_{f}$ & $890~\text{kg\,m}^{-3}$ \\
Surface tension & $\Gamma$ & $0.0309~\text{N\,m}^{-1}$ \\
Solvent viscosity & $\eta_{s}$ & $1.46~\text{Pa\,s}$ \\
Polymeric viscosity & $\eta_{p}$ & $13.14~\text{Pa\,s}$ \\
Relaxation time & $\lambda$ & $0.018225~\text{s}$ \\
Extensibility parameter & $\epsilon$ & $0$ \\
LPTT damping function & $f$ & $1$ \\
Normal stress difference ratio & $\psi_0$ & [0, 0.05, 0.15, 0.25] \\
\bottomrule
\end{tabular}
\label{tab:LPTTparams}
\end{table}

\begin{table}[h!]
\centering
\caption{Material parameters used for the Newtonian air phase.}
\begin{tabular}{l c c}
\toprule
\textbf{Parameter} & \textbf{Symbol} & \textbf{Value} \\
\midrule
Density & $\rho_{a}$ & $1.2~\text{kg\,m}^{-3}$ \\
Dynamic viscosity & $\eta_{a}$ & $1.0\times10^{-5}~\text{Pa\,s}$ \\
\bottomrule
\end{tabular}
\label{tab:airparams}
\end{table}

The computations assume an incompressible viscoelastic liquid with no-slip boundary conditions on all solid surfaces, including the rotating rod of radius $a=0.00635$ m. At the free surface, the kinematic condition is enforced together with the full normal and tangential stress balances, including capillary forces. The other phase is considered to be incompressible air with a constant density $\rho_a=1.2$ kg/m$^2$ and viscosity $\eta_a = 10^{-5}$ Pa.s. For simplicity and to eliminate the dependence on contact dynamics, and to focus only on the effects of $N_2$, we assume a contact angle of $90^{\circ}$ at the rod-liquid-air interface. This assumption is consistent with prior experiments and numerical studies \cite{beavers1975rotating, luo1999numerical, figueiredo2016two} and allows comparison with prior studies for validation. Furthermore, the Capillary number $Ca = (\eta_{s}+\eta_{p})\,\Omega a / \Gamma$, which characterises the relative magnitude of viscous to capillary forces, is $Ca \gg 1$ in all cases considered. This indicates that surface tension plays only a minor role in interfacial dynamics relative to viscous and elastic stresses. The interface is resolved using the volume-of-fluid approach implemented in InterFOAM \cite{deshpande2012evaluating}. At the far-field radial and axial boundaries, zero-normal-gradient conditions are applied for velocity, pressure and polymeric 
stresses. The boundary locations are placed sufficiently far from the rod to prevent any influence on the near-rod flow or on the evolution of the free surface.

\section{Results and Discussion}

\subsection{Climbing height variation with $\psi_0$}

\begin{figure}
    \centering
    \includegraphics[width=\linewidth]{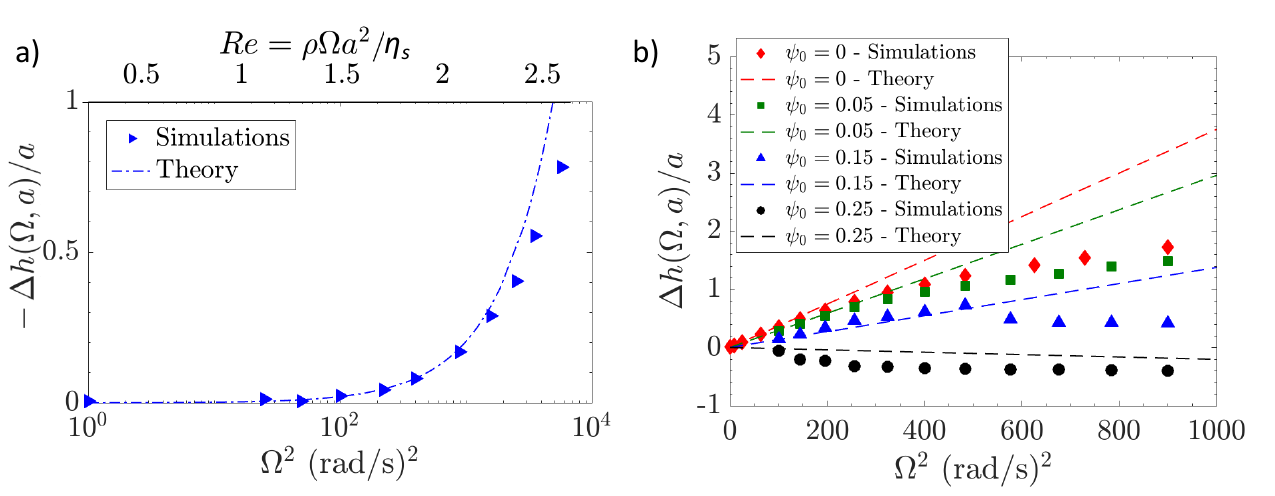}
    \caption{a) Comparison of the interface descending depth $-\Delta h(\Omega,a)$ scaled by the rod radius $a$ near the rod due to inertia for a Newtonian fluid from simulations (symbols) and theory predictions of Eq.~\ref{eq:eq2} (line). b) Effect of systematically increasing the normal stress difference ratio $\psi_0 = -\Psi_{2}/\Psi_{1}$ on the change in the interface height $\Delta h(\Omega,a)$ due to the Weissenberg effect, keeping all other fluid properties the same. Symbols are mean interface heights near the rod from the simulations at a long time, while dashed lines are predictions of Eq.~\ref{eq:eq2} using fluid properties in Table~\ref{tab:LPTTparams}. The agreement is good at low $\Omega$, and the simulation results diverge at high $\Omega$ as expected due to the validity of the theory only on the low rotation speed limit. The rod-climbing effect weakens as the second normal stress difference increases so much so that it transitions to rod-descending at the critical value $\psi_{0} \geq 0.25$, which gives climbing constant $\hat{\beta}=0$. However, note that the interface descends further than the perturbation theory predictions obtained using $\hat{\beta}=0$, which are valid only to $\approx O(\Omega^2)$.}
    \label{fig:fig3}
\end{figure}

It is well known that the interface descends due to inertial effects in a Newtonian liquid. This serves as a baseline for validating the solver's accuracy in this study. The Newtonian rod-descending height can be predicted from the perturbation solution in Eq.~\ref{eq:eq2} by substituting $\hat{\beta}=0$ as $\Psi_{1,0}=\Psi_{2,0}=0$ for Newtonian fluids. If we set $\eta_P=0$ Pa.s, $\lambda=0$ s and $\psi_0=0$, we recover an incompressible Newtonian fluid of constant viscosity $\eta_s = 1.46$ Pa.s and density $\rho_f = 890$ kg/m$^3$. The inertial rod-descending in the Newtonian limit is validated in Fig.~\ref{fig:fig3}a, confirming that the interface descends due to inertia with a magnitude consistent with predictions of Eq.~\ref{eq:eq2}. The simulation results diverge from the Eq.~\ref{eq:eq2} estimates at high rotation speeds, as the perturbation solution is not valid in that limit, as expected.

When viscoelasticity is incorporated into the fluid response, 
Fig.~\ref{fig:fig3}b shows that the climbing height decreases 
monotonically as the parameter $\psi_0$ is increased, while all other 
Rheological parameters are held fixed. The dashed curves, corresponding to the asymptotic prediction in Eq.~\ref{eq:eq2}, exhibit excellent agreement with the numerical results in the limit of small rotation rates. At higher angular velocities, however, the simulation results progressively deviate from these low-$\Omega$ theoretical estimates of Eq.~\ref{eq:eq2}.

This trend indicates that increasing the magnitude of the second normal 
stress difference $N_2$ systematically suppresses the rod-climbing effect, ultimately yielding progressively smaller interface elevations. Beyond a threshold value, specifically around $\psi_0 \approx 0.25$, the climbing height not only vanishes but reverses sign---the free surface is drawn \emph{downward} rather than upward. In this regime, the system exhibits a distinctly \emph{viscoelastic rod-descending} response. The dimensionless ratio $\rho a^{2}/\Psi_{1,0} \approx 0.1$ indicates that elastic stresses dominate over inertial contributions in this regime. Consequently, inertia-driven rod-descending can be ruled out as a possible mechanism \cite{more2023rod}.

These findings reveal that a viscoelastic fluid can exhibit rod-descending behaviour even in the absence of significant inertia and despite a substantial positive first normal stress difference, $N_1$. Consequently, the classical Weissenberg rod-climbing effect cannot be regarded as an unequivocal diagnostic of viscoelasticity under all circumstances. Instead, these results highlight an important exception to the conventional expectation that rod climbing is a universal signature of viscoelastic normal-stress effects.

\subsection{Competition between $N_1$ and $N_2$ governs the weakening and reversal of rod climbing}

\begin{figure}
    \centering
    \includegraphics[width=\linewidth]{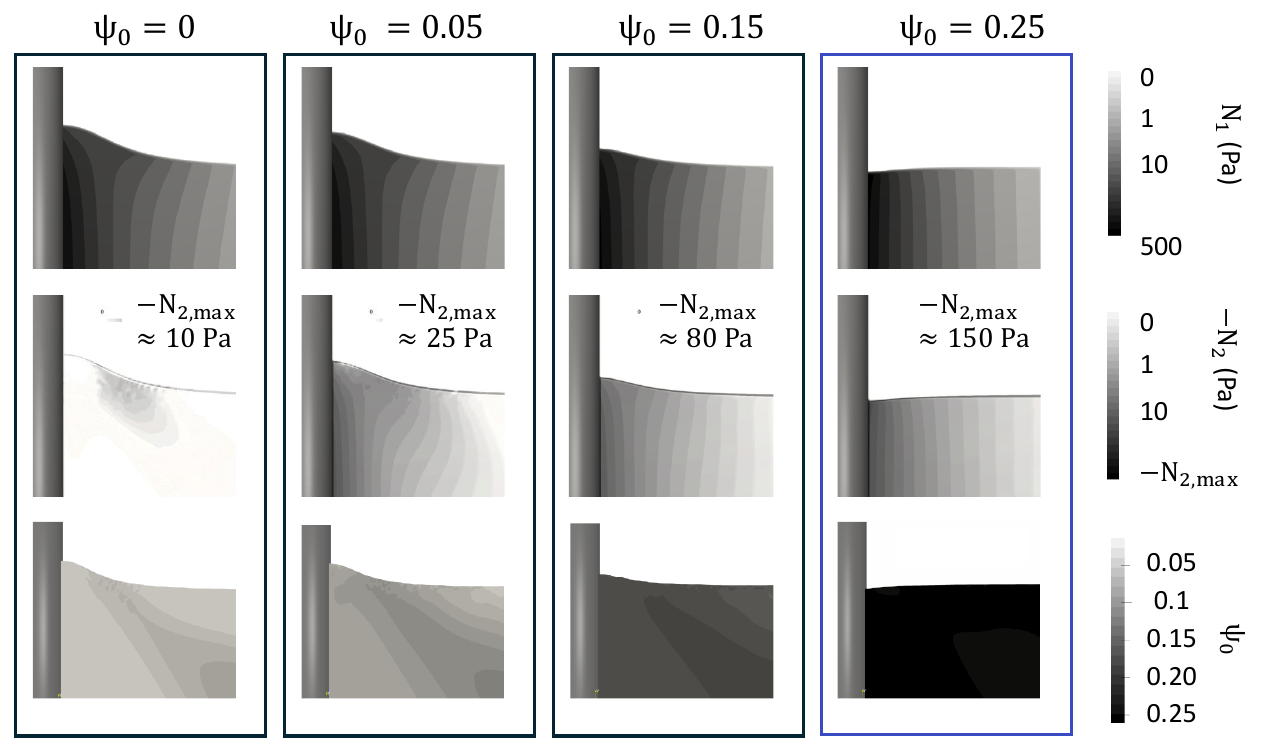}
    \caption{Contours of the first (top row), second (middle row) and the normal stress difference ratio (bottom row) for increasing values of the second normal stress difference relative to the first normal stress difference (left to right columns). The first normal stress difference has almost a similar magnitude near the rod as well as a similar contour, irrespective of $\psi_0$. However, we observe weakening rod-climbing with increasing $\psi_0$, which can be explained by the increasing magnitude of the contours of $N_2$ with increasing $\psi_0$. So, while $N_1$ pushes the liquid up, $N_2$ works in opposition to it and is dominant enough to reverse the rod-climbing to rod-descending beyond a critical value $\psi_0=0.25$. The governing role of $\psi_0$ is also clear from its contours in the bottom row, which shows progressively increasing and almost uniform $\psi_0$ field in the entire liquid.}
    \label{fig:fig4}
\end{figure}

The physical mechanism underlying the progressive weakening and eventual reversal of the rod-climbing response with increasing $\psi_{0}$ becomes clear upon examining the normal stress difference fields in Fig.~\ref{fig:fig4}. The figure shows contours of $N_{1}$ (top row), $N_{2}$ (middle row), and the normal-stress ratio $\psi_{0}$ (bottom row) for increasing $\psi_{0}$ at a fixed rotation rate of $\Omega = 10~\text{rad}\,\text{s}^{-1}$ when steady state is achieved. Across all simulations, the first normal stress difference $N_{1}$ remains essentially unchanged as $\psi_{0}$ is increased from $0$ to $0.25$, indicating that the elasticity associated with streamwise stretching is largely insensitive to variations in $N_{2}$. Based on the $N_{1}$ field alone, one would therefore expect a monotonic rod-climbing response as popularly believed. In contrast, the computed interface exhibits progressively smaller climbing heights and ultimately transitions to rod‐descending at $\psi_{0}=0.25$. 

This behaviour originates from the systematic and substantial increase in the magnitude of the second normal stress difference $N_{2}$ with increasing $\psi_{0}$. Whereas $N_{1}$ generates the familiar hoop stresses that drive fluid upward along the rod, $N_{2}$ introduces a meridional compressive stress that acts to pull the interface downward. As $\psi_{0}$ increases, this compressive contribution intensifies and competes directly with the upward contribution from $N_{1}$. Once $|N_{2}|$ becomes sufficiently large, the meridional compression dominates, reversing the direction of free-surface deformation and leading to the rod descending.

As $\psi_0$ becomes sufficiently large, the compressive stresses linked to $N_2$ grow strong enough to dominate the normal-stress balance, ultimately overwhelming the lifting effect of $N_1$. When this occurs, the net normal stress acting on the free surface reverses sign, causing the interface to dip instead of rise and thereby producing viscoelastic rod-descending rather than rod-climbing. Importantly, this transition does not arise from a localised perturbation of the stress field in the high-shear zone adjacent to the rotating rod. Rather, the nearly uniform distribution of $\psi_0$ throughout the domain indicates that the reversal is a global force-balance phenomenon. The entire viscoelastic medium experiences a modified distribution of normal stresses, meaning that the climbing reversal is governed by a domain-wide redistribution of meridional compressive stresses rather than a boundary-layer or near-rod effect.

This global character of the stress redistribution distinguishes the reversal mechanism from classical interpretations of the Weissenberg effect, which typically emphasise localised hoop and tensile stresses near the rod surface \cite{bird1987dynamics}. Instead, the present results demonstrate that sufficiently significant negative contributions from $N_2$ can reorganise the overall stress balance of the fluid to such an extent that the canonical rod-climbing behaviour is negated and replaced by its inverse.

\subsection{Rod-descending at $\psi_0 \gtrapprox 0.25$ arises from elasticity, not inertia}

\begin{figure}
    \centering
    \includegraphics[width=0.8\linewidth]{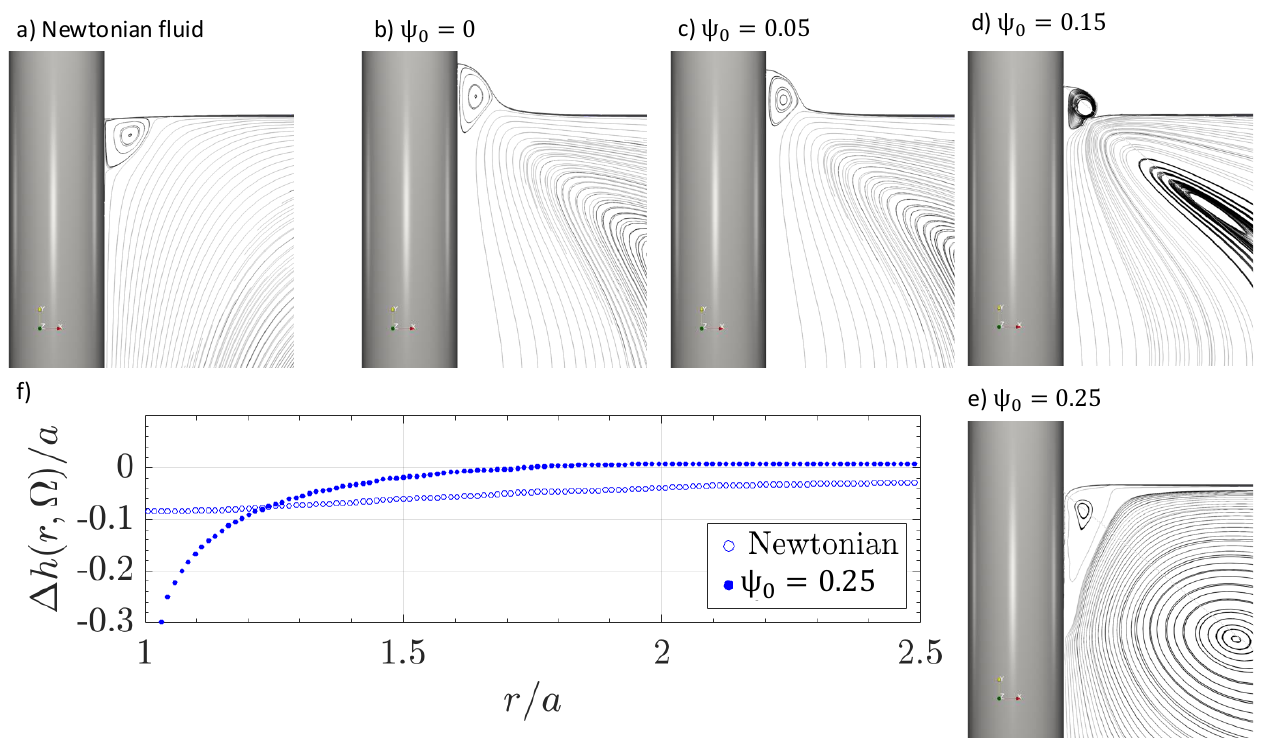}
    \caption{Streamlines around the rotating rod for Newtonian and viscoelastic liquids. (a) In a Newtonian fluid, only a single primary recirculation cell forms. (b–e) In contrast, viscoelastic simulations exhibit a distinct secondary vortex generated by normal-stress differences, consistent with experimental observations \cite{saville1969secondary}. The strength and extent of this secondary vortex increase systematically with the magnitude of the second normal stress difference $N_{2}$ or $\psi_0$ [(b) $\psi_0=0$, (c) $\psi_0=0.05$, (d) $\psi_0=0.15$, (e) $\psi_0=0.25$], analogous to behaviour reported in confined viscoelastic flows through non-circular geometries \cite{maklad2021review}. (f) A comparison between rod-descending in a Newtonian fluid (driven purely by inertia) and in a viscoelastic fluid with $\psi_{0}=0.25$ (corresponding to $\hat{\beta}=0$) highlights the role of higher-order $O(\Omega^{4})$ contributions that are neglected in the perturbation expression given in Eq.~\ref{eq:eq2}.}
    \label{fig:fig5}
\end{figure}

The flow fields at $\Omega = 20$ rad/s for varying $\psi_0$ in the form of streamlines are presented in Fig.~\ref{fig:fig5}, providing additional insight into the mechanisms responsible for the reversal of the rod-climbing response. For a Newtonian fluid ($Re=0.5$), the meridional circulation consists solely of a single, well-defined primary vortex. In contrast, viscoelastic fluids ($Wi=0.36$, $Re=0.05$ and $\rho a^2/\Psi_{1,0} = 0.11$) develop a qualitatively different flow topology: even when the magnitude of the second normal stress difference $N_2$ is modest ($\psi_0 = 0.05$), a distinct secondary vortex forms in the vicinity of the rod. This feature is absent in Newtonian flows and is consistent with experimentally observed vortices near the rotating rod in a viscoelastic fluid \cite{saville1969secondary}.

As the normal stress difference ratio $\psi_0$ is increased further, the secondary vortex strengthens and expands, progressively restructuring the meridional circulation. The intensified vortex redistributes momentum, thereby enhancing downward surface curvature and facilitating the onset of rod-descending. This behaviour is not unique to the present configuration: the amplification of secondary motions by $N_2$ mirrors the mechanisms documented in other flow geometries, such as ducts, curved channels, and eccentric annuli, where the second normal stress difference is known to suppress or reverse classical viscoelastic secondary flows, as reviewed comprehensively by Poole and Maklad (2021)~\cite{maklad2021review}. Thus, the observed vortex restructuring represents a canonical manifestation of how $N_2$ modifies cross-stream transport in complex viscoelastic flows.

The emergence and intensification of the secondary vortex occur concurrently with a shift in the normal-stress balance: the downward-directed contribution from $N_2$ increasingly offsets, and ultimately nullifies, the upward hoop stress produced by $N_1$ as discussed in the last sub-section Sec.IV-B. Notably, the magnitude of the interface dipping observed in the simulations far exceeds the second-order perturbation prediction obtained by setting $\hat{\beta}=0$ for $\psi_0 = 0.25$ in Eq.~\ref{eq:eq2}, as illustrated in Fig.~\ref{fig:fig5}f. This discrepancy arises because the higher-order nonlinear contributions \cite{beavers1975rotating}, specifically the $O(\Omega^4)$ terms in Eq.~\ref{eq:eq2}, no longer remain negligible in the vicinity of $\psi_0 = 0.25$. Thus, viscoelasticity can produce a rod-descending response whose magnitude 
far exceeds that of the inertial rod-descending observed in a Newtonian fluid of the same density $\rho$.

This behaviour is fully consistent with the classical observations of Beavers and Joseph~\cite{beavers1975rotating, beavers1980free3}, who showed that perturbative theories can substantially underpredict free-surface deformation when nonlinear viscoelastic effects become comparable in magnitude to the leading-order normal stresses. Collectively, these results highlight the necessity of retaining the full nonlinear form of the constitutive model and $O(\Omega^4)$ terms when analysing rod-climbing and rod-descending responses in highly elastic fluids. A comprehensive nonlinear asymptotic treatment, however, lies beyond the scope of the present study.

\subsection{Beyond low rotation speeds: onset of unsteadiness and chaos}

\begin{figure}[h!]
    \centering
    \includegraphics[width=0.85\linewidth]{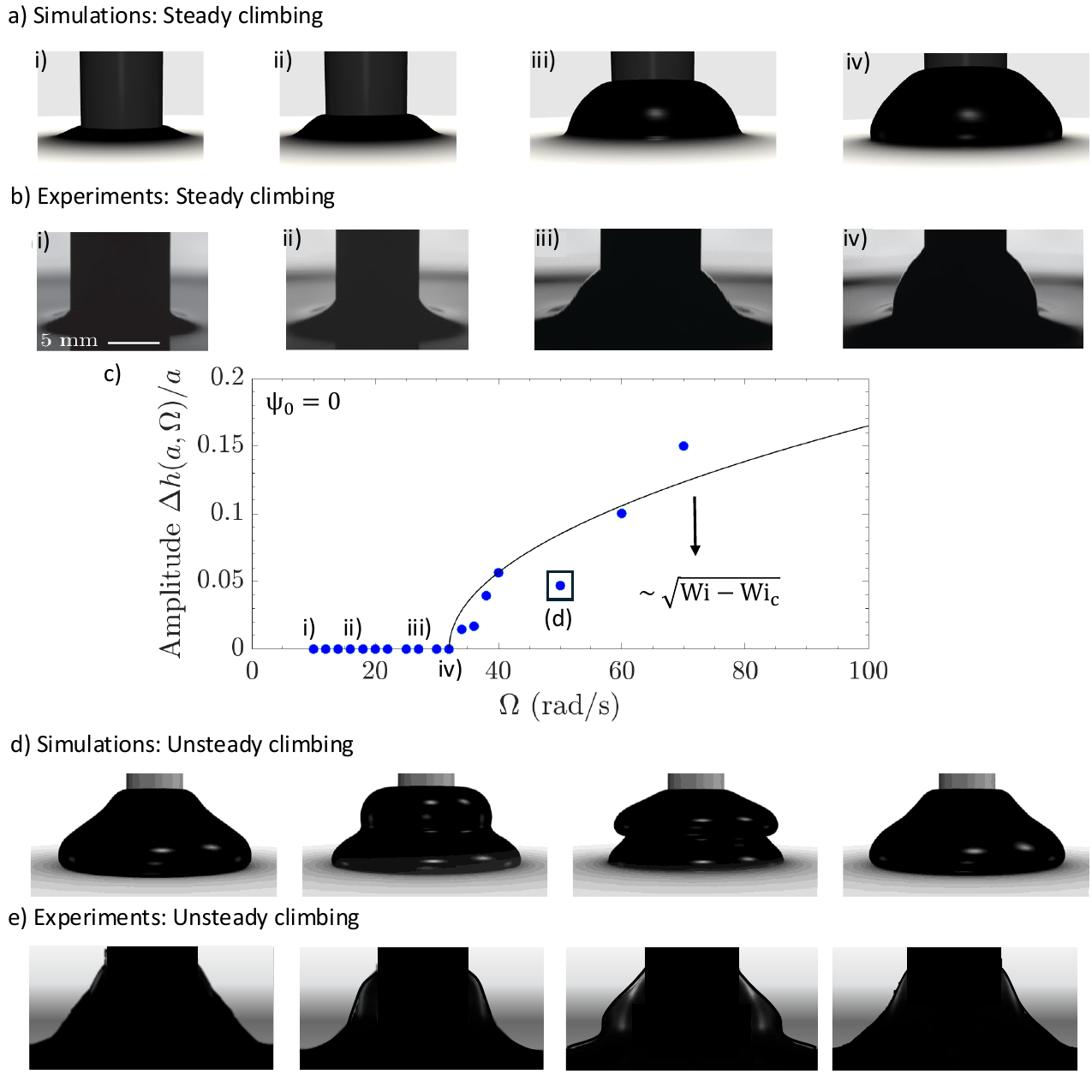}
    \caption{Simulation snapshots (a, d) illustrate the sequence of transitions observed as the rod rotation rate increases, with corresponding experimental images for a 3 wt.\,\% PIB solution~\cite{more2023rod} shown in (b, e) for qualitative comparison. Panels (a)–(b) show the steady climbing regime. Panels (i)–(ii) ($\Omega=10$–20 rad/s) correspond to regime~(I), where only small steady perturbations appear. Panels (iii)–(iv) ($\Omega=30$–40 rad/s) show regime~(II), the steady bubble state in which accumulated fluid near the rod eventually rolls downward. The oscillation amplitude at $\Omega=50$ rad/s, plotted in (c), reflects a subcritical Hopf bifurcation. Panels (d)–(e) illustrate regime~(III), characterised by unsteady oscillations and periodic rollover of fluid sheets. A supplementary video shows the oscillatory behaviour in both simulations and experiments.}
    \label{fig:fig6}
\end{figure}

At higher Weissenberg numbers, the system recovers the complete sequence of nonlinear transitions documented experimentally \cite{beavers1975rotating, degen1998time, more2023rod}. Fig.~\ref{fig:fig6}a-i, ii (present simulations) and Figure~\ref{fig:fig6}b-i, ii (earlier experiments \cite{more2023rod} of 3 wt.\,\% PIB in oil polymer solution for comparison) demonstrate that at low $\Omega$ the interface remains in a steady climbing state, consistent with the classical viscoelastic response. As $\Omega$ increases, a localised bubble-like elevation forms adjacent to the rod, marking the onset of a strongly nonlinear regime (Figure~\ref{fig:fig6}a-iii and b-iii). As this climbing bubble grows (Figure~\ref{fig:fig6}a-iv and b-iv), the elastic stresses that support it eventually become insufficient to counteract the hydrostatic load. The bubble then collapses, producing sustained periodic oscillations of the free surface.

The amplitude of these oscillations with increasing $\Omega$ is plotted in Fig.~\ref{fig:fig6}c, follows a subcritical Hopf bifurcation, in agreement with the dynamical behaviour identified in earlier rotating-rod experiments \cite{degen1998time}. The oscillating interfaces in a single oscillation cycle at $\Omega = 50$ rad/s are depicted in Fig.~\ref{fig:fig6}d and e for simulations and experiments \cite{more2023rod}, respectively. At even higher rotation rates, the oscillations progressively lose symmetry, exhibiting waveform distortion, temporal intermittency, and ultimately rupture of the free surface. This progression accurately reproduces the transitions reported in unsteady free-surface studies of viscoelastic rod climbing \cite{beavers1975rotating,degen1998time,more2023rod}, thereby demonstrating that the present model captures the experimentally observed hierarchy of nonlinear states. 

It should be noted, however, that because the present simulations are 2D axisymmetric, they cannot faithfully represent the fully three-dimensional nature of the highest Weissenberg-number regimes. As the flow approaches these strongly nonlinear states, axisymmetric computations become numerically unstable, signalling the onset of three-dimensional physical modes. Nonetheless, the simulations still exhibit clear precursors of this transition: the loss of symmetry, irregular oscillations, and eventual free-surface rupture observed in our results can be interpreted as the axisymmetric manifestation of the 3D flow reorganisation that occurs experimentally. A supplementary video illustrates this behaviour for $\psi_{0}=0$ at $\Omega = 100~\text{rad}\,\text{s}^{-1}$, where the axisymmetric solution visibly develops features characteristic of a transition to fully three-dimensional rod-climbing dynamics.

\subsection{$\psi_0$ vs. $Wi$ phase diagram of rod-climbing transitions}

\begin{figure}
    \centering
    \includegraphics[width=0.6\linewidth]{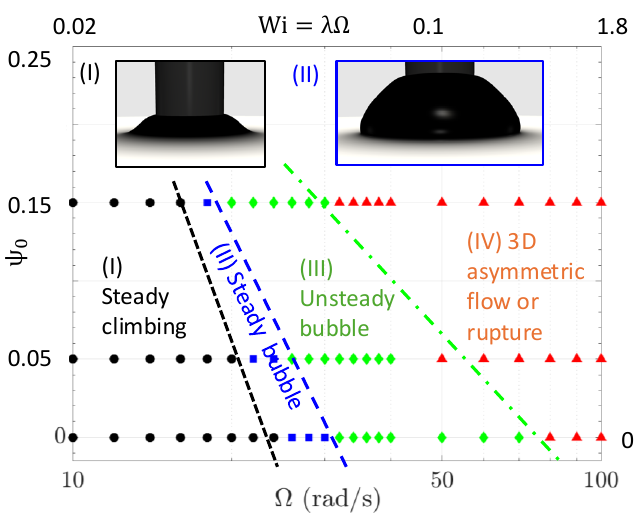}
\caption{Phase diagram in the $(\psi_{0},Wi)$ space illustrating the sequence of transitions observed in the rod-climbing system: (I) steady climbing, (II) steady bubble formation, (III) oscillatory or unsteady bubble dynamics, and (IV) asymmetric three-dimensional flow leading to rupture. These regimes are encountered as the rod rotation rate $\Omega$ (and hence $Wi$) is increased. As shown in Fig.~\ref{fig:fig4} and Fig.~\ref{fig:fig5}, increasing the magnitude of the second normal stress difference progressively weakens the rod-climbing response and simultaneously lowers the thresholds for the onset of unsteady behaviour. Beyond the critical value $\psi_{0}\approx 0.25$, the system no longer supports climbing and instead transitions to rod-descending.}
    \label{fig:fig7}
\end{figure}

Finally, simulation results for a range of $\psi_0$ and $\Omega$ enable a quantitative determination of the critical rotation rate $\Omega$ associated with each transition, namely, from small steady perturbations to steady bubble formation, to oscillatory bubbling, to fully three-dimensional asymmetric flow, and eventually to rupture. This systematic quantification across the $(Wi,\psi_0)$ parameter space is presented in Fig.~\ref{fig:fig7}, yielding a comprehensive phase diagram with distinct regimes, as illustrated in Fig.~\ref{fig:fig6}. The effect of increasing $N_2$ is clear: it makes the rotating flow of a viscoelastic fluid increasingly unstable as the transitions occur at lower and lower $\Omega$ with increasing $\psi_0$. Finally, beyond $\psi_0 \approx 0.25$, steady climbing is no longer observed: the system undergoes either immediate rod-descending or a direct transition into unsteady, asymmetric surface dynamics.

\section{Conclusions}\label{sec:sec5}

This work has demonstrated that the second normal stress difference $N_2$ plays a crucial and previously underappreciated role in the free-surface deformation and stability of rod-climbing flows. Using fully resolved axisymmetric simulations of the LPTT model, with the normal-stress ratio 
$\psi_0 = -N_2/N_1$ treated as an independent control parameter, we have shown that increasing $|N_2|$ systematically weakens the Weissenberg effect and ultimately reverses the sign of the interface height at the rod. Beyond a critical value $\psi_0 \approx 0.25$, the classical rod-climbing response is replaced by a purely viscoelastic rod-descending state, even in the absence of significant inertia and in the presence of a substantial positive $N_1$.

The simulations further reveal that $N_2$ modifies the internal flow topology and stress distribution in a manner that strongly promotes instability. As $\psi_0$ increases, the secondary vortex near the rod is amplified, the meridional normal-stress field is reorganised, and the onset of nonlinear transitions---steady bubble formation, subcritical Hopf oscillations, three-dimensional asymmetric flow, and free-surface rupture---shifts to progressively lower Weissenberg numbers. The resulting $(Wi,\psi_0)$ phase diagram consolidates these regimes into a unified framework. It introduces the normal-stress difference ratio $\psi_0$ as a third fundamental dimension, alongside the Weissenber $Wi$ and Reynolds $Re$ numbers, for understanding complex-fluid 
behaviour and the onset of viscoelastic flow instabilities.

Findings from this study show that free-surface viscoelastic flows are not governed solely by the magnitude of the first normal stress difference, but by a delicate competition between $N_1$ and $N_2$. This competition controls both the sign and the stability of the interface deformation, in close analogy with how elasticity and inertia jointly structure bulk-flow instabilities in complex fluids \cite{more2023rod, more2024elasto}. The present results help reconcile discrepancies between classical perturbation theory, rotating-rod experiments, and simulations by explicitly identifying $N_2$ as a key rheological parameter.

More broadly, this study demonstrates that rod climbing should be regarded not merely as a qualitative indicator of the presence of normal stresses but as a sensitive probe of their relative magnitudes and dynamical interplay. This has direct implications for the interpretation of rod-climbing viscometry \cite{beavers1975rotating, more2023rod}, the rheological characterisation of polymer solutions and industrial fluids, and the prediction and control of free-surface instabilities in processing flows \cite{maklad2021review} where both $N_1$
and $N_2$ are non-negligible.

\section*{Supplementary Materials}

The supplementary materials include three videos illustrating the unsteady dynamics observed in the rotating rod experiment. The first video (Title: Unsteady oscillations in rod-climbing (experiments)) shows unsteady interface oscillations at a rotation speed of $50~\text{rad\,s}^{-1}$ in a 3~wt.\,\% polyisobutylene solution, highlighting the periodic rollover and amplitude modulation characteristic of regime~(III) oscillations. The second video (Title: Unsteady oscillations in rod-climbing (simulations)) presents the corresponding behaviour in the present axisymmetric simulations for $\psi_{0}=0$, demonstrating that the numerical method reproduces the essential oscillatory dynamics observed experimentally. The third video (Title: Rupture at high rotation speeds in rod-climbing) captures the manifestation of onset of three-dimensional flow and the eventual rupture of the fluid column at very high rotation speeds ($100~\text{rad\,s}^{-1}$) for $\psi_{0}=0.05$ in a 2D flow, where the free surface loses symmetry and breaks apart once the accumulated mass can no longer be supported by the elastic stresses.

\begin{acknowledgements}

The author gratefully acknowledges Prof.\ Gareth McKinley for insightful discussions and for support with the experimental work. The author also thanks Lubrizol Inc.\ for providing the viscoelastic fluids used in this study. This research was supported in part by the Engineering Talent Research Accelerator Fellowship (2025) from the Faculty of Engineering at Monash University. Computational resources were provided by the Monash Massive High-Performance Computing facility.

\end{acknowledgements}

\section*{Data Availability Statement}

The data that support the findings of this study are available
within the article [and its supplementary material].

\appendix

\begin{figure}[h!]
    \centering
    \includegraphics[width=\linewidth]{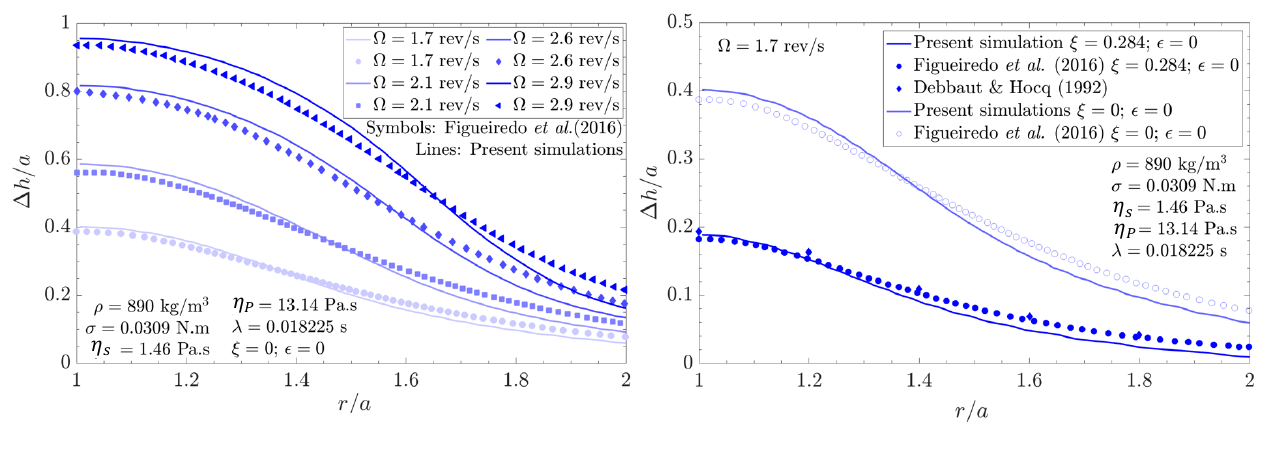}
\caption{Validation of the numerical methodology.  
(a) Comparison of steady free-surface profiles obtained using the present LPTT-based formulation with the results of Figueiredo \textit{et al.}~\cite{figueiredo2016two} for an Oldroyd--B (Boger) fluid, reproduced by setting $\psi_{0}=0$ and $\epsilon=0$ in the LPTT model. Interface shapes are shown at four different rotational speeds, demonstrating excellent agreement over the entire meniscus.  
(b) Comparison between the present simulations and the numerical results of Debbaut \textit{et al.}~\cite{debbaut1992numerical} for a viscoelastic fluid with normal-stress ratio $2\psi_{0}=\xi=0.284$. The close match confirms the computational framework's ability to reproduce benchmark rod-climbing profiles across distinct constitutive behaviours accurately.}
    \label{fig:fig8}
\end{figure}

\section{Validation of the numerical method}

Figure~\ref{fig:fig8}a compares the interface profiles obtained from the present simulations with those reported by Figueiredo \textit{et al.}~\cite{figueiredo2016two} for an Oldroyd--B fluid, which represents a constant-viscosity elastic (Boger) fluid. This correspondence is achieved within the LPTT framework by setting $\psi_{0}=0$ and $\epsilon=0$, as described in Sec .~II. Four rotational speeds are shown, and in all cases, the present results accurately reproduce the previously reported free-surface shapes. For clarity, we note that the parameter used in \cite{figueiredo2016two}, denoted $\xi$, relates to our notation through $\xi = 2\psi_{0}$. 

Figure~\ref{fig:fig8}b further compares the interface profiles predicted by the current methodology with those from Debbaut \textit{et al.}~\cite{debbaut1992numerical} for a case with $2\psi_{0}=\xi = 0.284$. The agreement across the entire meniscus is excellent, thereby validating the accuracy and robustness of the numerical approach employed to simulate the rod-climbing problem.

\bibliography{aipsamp}

\end{document}